\documentclass{iopart}
\usepackage{amssymb}
\usepackage{graphicx}
\usepackage{bm}
\usepackage{dcolumn}
\usepackage{longtable}
\begin{document}
\title[Teaching Physics using video games]{Teaching introductory undergraduate Physics using commercial video games}
\author{Soumya D.~Mohanty and Sergio Cantu}
\address{Department of Physics and Astronomy, The University of Texas at Brownsville, 80 Fort Brown, Brownsville, TX 78520, USA.} 
\ead{mohanty@phys.utb.edu}
\begin{abstract}
Commercial video games are increasingly using sophisticated physics simulations to create a more immersive experience for players. This also makes them a  powerful tool for engaging students in learning physics. We provide some examples to show how commercial off-the-shelf games can be used to teach specific topics in introductory undergraduate physics. The examples are selected from a course taught predominantly through the medium of commercial video games.
 \end{abstract}
\section{ Introduction} 
One of the key determinants of the commercial success of a video game is the degree of
 immersive experience provided to players.  To achieve this, game developers are increasingly 
 relying on physics based virtual worlds, where the consequences of an action are not pre-programmed but
  computed in real time using some physical laws.  
  
  Specialized software libraries, called Physics engines, 
   allow physics to be easily integrated with other elements to create 
 realistic games. 
 A study of 
 several commercial and free Physics engines (Boeing \& Brunl, 2007) shows that the accuracy of
physics simulation is already quite good. Ongoing hardware improvements will only make them better.

Given the preponderance of physics in commercial video games, it is natural to ask if they
   can be parlayed into teaching physics effectively. 
 

To answer this question
we put commercial games to the test by using them as the main medium 
 for teaching an entire physics course. The course, offered
 during Dec, 2010, to Jan, 2011, at the University of Texas at Brownsville (UTB), was at the
introductory undergraduate level. 
We present selected examples from this course to show how commercial games 
can be put to good use in physics teaching.

Some of the games used in the course are briefly described below. All of them are playable on the Playstation-3 (PS3) console.  
\begin{itemize}
 \item Little Big Planet (LBP): An environment for  creating games and sharing them over the internet, this meta-game
has a library of tools and objects for building virtual machines.
 The simulation of Newtonian dynamics in such constructions 
 is quite impressive in its scope and fidelity, but it is limited to an
 essentially two-dimensional world. We note that LBP could meet the requirements of 
 an ``immersive environment" for learning physics as defined by Price (2008).
\item  Shaun White Skateboarding (SWS): A game with a simple narrative in which the player controls a skateboarder. The graphics in this game
are low-key but, from the perspective of teaching, provide good support for introductory kinematics.
\item  Guardian of light (GLight): The latest installment in the popular {\it Tomb Raider} franchise, this game features
 cooperative play by up to 2 players. This allows setting up in-game collaborative physics experiments. 
\item Uncharted 2 (UnC2): The second installment in the {\it Uncharted} franchise, this hugely popular game pushes the PS3 to the extreme. This game has top of the line visuals, a deep and engaging story and a significant amount of physics. It
 is cumbersome to set up repeatable experiments in this game but it provides good opportunities for observing ``real-world"  physics while engaging students. 
\end{itemize}
More detailed descriptions of these games can be obtained from the web. Several of them have {\em Wikipedia} articles and game play videos on {\em Youtube}. 

The rest of the paper is organized as follows. Sec.~\ref{examples_kinematics} contains examples related to the 
topic of kinematics in Newtonian mechanics and Sec~\ref{examples_dynamics} has examples from dynamics.  The examples are in the form of a set of
activities followed by an outline of  the pedagogical motivation behind them. 
Some 
general observations based on our experience with the UTB course 
are discussed in Sec~\ref{discussion}. 
\section{Examples related to Kinematics}
\label{examples_kinematics}

\subsection{Coordinates }
\label{coordinates}
\begin{figure}
\includegraphics[scale=0.3,angle=-90]{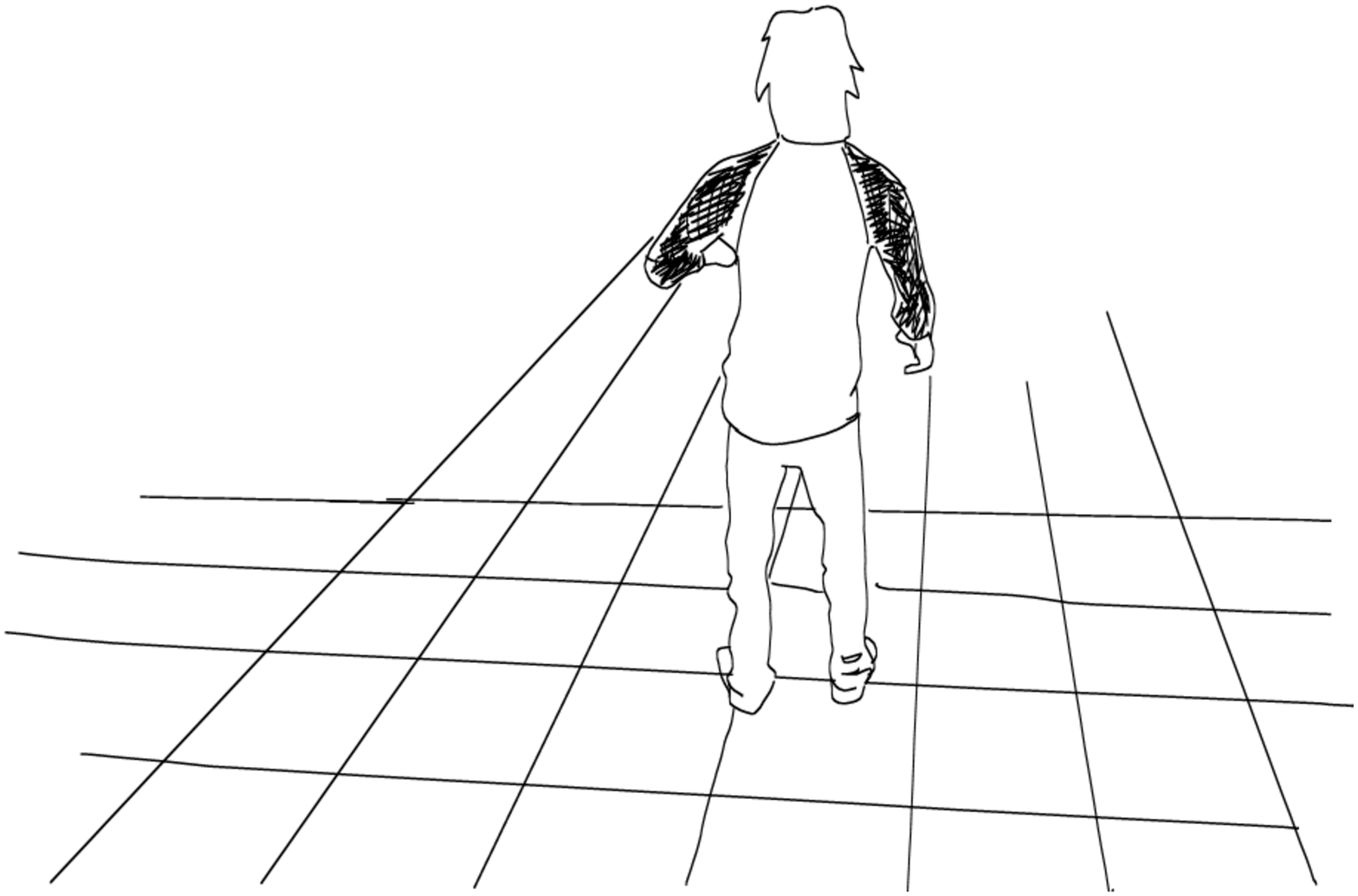}
\caption{\label{figure1}
A two dimensional Cartesian coordinate system using a tiled floor in the SWS game.
}
\end{figure}

 Fig.~\ref{figure1} shows an image from the SWS game where the controllable character is standing on a floor that has a regular grid of tiles. Students can use the tiles as a unit of length and carry out several exercises. 
 
\noindent {\it Activities:}
\begin{enumerate}
\item Students construct their own two-dimensional Cartesian coordinate system by picking an origin on the floor and XY axes along two mutually perpendicular grid lines on the floor. They then measure the coordinates of various objects on the floor by walking the character along the axes directions and measuring the distances along each in ``tile" units.  
\item  Pick axes directions at $45^\circ$ to the grid lines and redo the measurements.
\item Students calculate the Euclidean distance between two points using the difference between their coordinates. Using different coordinate labels for the same two locations, they verify that the distance between them is invariant.
\end{enumerate}
The first activity teaches both the concept of a measurement unit, length in this case, and the concept of identifying locations using numerical ``address labels" (coordinates). The second activity clarifies the difference between the physical nature of a quantity and its observer dependent mathematical description. The third activity reinforces this concept for the case of distance between two points.  

The activities can be extended to the third dimension by using a level, not shown here, that has stairs in it. The stairs provide a different unit of length in the vertical direction, thus establishing the concept that coordinates need not have the same units or even the same physical dimensions.

\subsection{Basic measurements and errors}
\noindent {\it Activity:}
Students measure the speed of the character  in SWS when walking, running or skateboarding by using a stopwatch to measure the time required to cover a fixed distance (in tile units). 

In SWS, as in most other games, motion comes in limited variety and the corresponding speeds are fixed. Thus, the error in measurement comes only from sources that students can appreciate readily, such as the resolution limit imposed by the length unit or errors in timing. In contrast, real systems have extraneous effects such as friction induced acceleration or start and end transients, that could confuse and frustrate the beginner. 

Measurement activities in games cannot replace 
 laboratory experiments. However, for the reasons mentioned above, they can
serve as a good hands-on introduction to the concepts that students need to master in actual experiments. 


\subsection{Paths and introduction to kinematics}
\label{paths}
\begin{figure}
\includegraphics[scale=0.3,angle=-90]{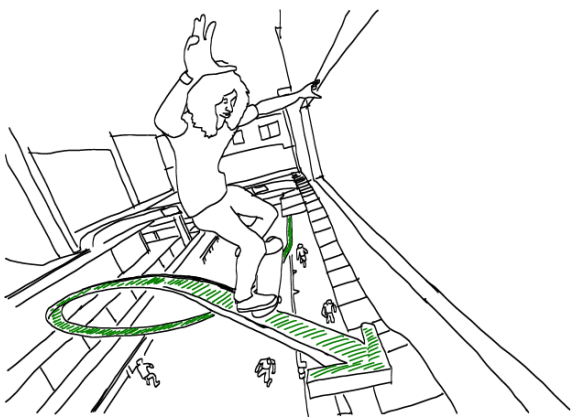}
\caption{\label{figure2}
Visualization of paths (green curve) in three-dimensional space in the SWS game. The arrow points along the instantaneous direction of travel.
}
\end{figure}

Fig.~\ref{figure2} shows an interesting aspect of the SWS game: the main character can ride special grind rails that dynamically evolve on the screen into three-dimensional paths. Once constructed, these tracks persist in the game. 

\noindent {\it Activities:}
\begin{enumerate}
\item 
Observe how a path evolves in three-dimensional space and once a path has been created, observe it from the ground or other vantage points reachable by the character using various camera angles. (The camera can be controlled by the player.)
\item Estimate and assign coordinates to points along the path and note the times at which the player crosses those points. The times can differ from one trial to another for the same path depending on the initial speed and the need to maintain balance while on the path.
\item Understand the difference between the trials as the outcome of different speeds on the same spatial trajectory.
\end{enumerate}
The first activity explores the concept of paths in three-dimensional space and the physical, observer independent, nature of such paths. The second activity introduces the concept of coordinates of a moving particle as functions of time. Since a given point on a curve can be traversed at different times, the concepts of instantaneous and average speeds along a path are introduced in the
third activity. The evolving path in the game has an arrowhead that points in the instantaneous direction of travel. Coupling this with  instantaneous speed serves to introduce the instantaneous velocity vector. All this can be done without explicit reference to calculus.

\subsection{Visualizing vector quantities}

\begin{figure}
\includegraphics[scale=0.3,angle=-90]{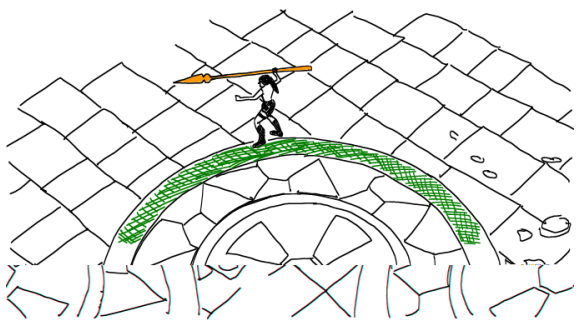}
\caption{
\label{velocityvector}
The spear in the Glight game as a representation of the velocity vector. As the player moves the character along the circle, the spear must be simultaneously controlled to remain along the tangent.
}
\end{figure}

In the Glight game, one of the characters can be equipped with a spear. When so equipped, the spear is held as shown in 
Fig.~\ref{velocityvector}
and it turns out that the speed of the character becomes constant (walking speed). The spear, therefore, offers a ready visual representation of a velocity vector for the case of constant speed. The direction of the spear can be changed independently by the player with respect to the direction of motion.

\noindent {\it Activity:}
Make the character walk on prescribed paths while keeping the velocity vector (the spear) pointing in
 the correct direction, that is tangential to the path. For example, the path could be a 
 circle as shown in Fig.~\ref{velocityvector} and the student must keep reorienting the spear while making the 
character moving on the circle.

This activity reinforces
 the concept, introduced with the SWS game, that a velocity vector is always tangential to the path.  Controlling the spear properly on a circular path reinforces the fact that a vector quantity can change without a change in magnitude.  
By putting the student in control of a vector,  the physical nature of an abstract mathematical concept is elucidated.


\section{Examples related to Dynamics}
\label{examples_dynamics}

\subsection{The nature of physics}
Once the student has understood the idea of the state of a particle and its mathematical description (kinematics), 
the concept can be introduced that all of physics is about connecting the current or future state of a system to a previous one. 
The connecting element is a set of physical laws, be it those of Newtonian or Quantum Mechanics, or the ones designed for a game. 

\noindent {\it Activity:} The student is asked to ponder over what the processor in a game console is computing between one frame on the screen to the next. The goal is to understand that it is connecting the previous state of a system to the next one using some ``Law". This provides a concrete illustration of how nature itself evolves systems following some universal laws.

This activity sets the stage for the introduction of dynamics and Newton's laws. The discussion can also touch upon other types of physical laws, such as the Maxwell equations for electro-magnetic phenomena. 

\subsection{Newton's laws}
\begin{figure}
\includegraphics[scale=0.5,angle=-90]{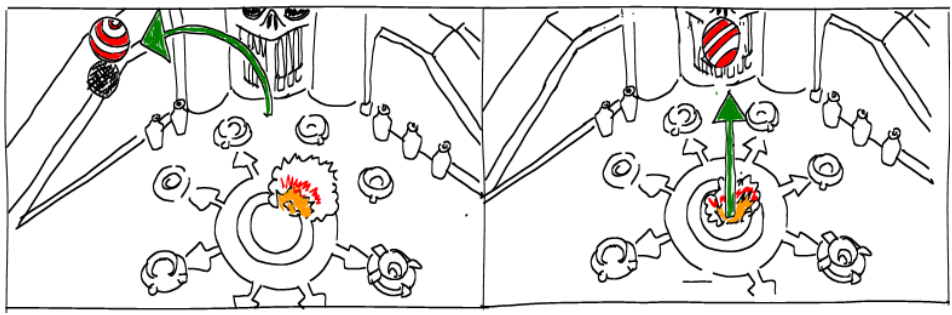}
\caption{
\label{bombsaway}
The right panel shows the effect of placing bombs on one side of a sphere in the Glight game. The same two bombs placed on opposite sides lead to a different trajectory as seen in the left panel. The trajectories are indicated schematically by the green arrows.
}
\end{figure}

\begin{figure}
\includegraphics[scale=0.3,angle=-90]{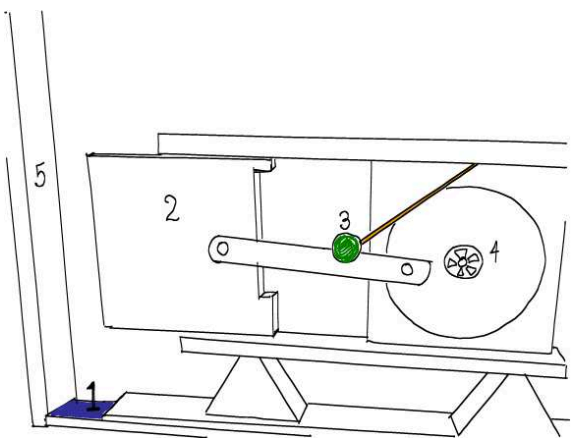}
\caption{\label{lbpsled}
Understanding the third law in LBP. The various parts of the machine are as follows.
 (1) The frictionless track on which the sled moves. (2)
The piston head that pushes against the wall. (3) The bob of the pendulum used to show acceleration. (4)  The motor object which 
moves the piston head. (5) The wall at the end of the track.
}
\end{figure}

 Glight and LBP  can be used to explore the consequences of Newton's laws as follows. 
\noindent {\it Activities:}
\begin{enumerate}
\item The LBP game provides a choice of materials and one of them, called ``glass", is supposed to be frictionless. Students construct long tracks of different materials and use a  piston to kick an object placed on the track. They observe that in the absence of friction an object will move with uniform motion. This activity could be used as a prelude to a real lecture demonstration based on airtracks to show students that the first law in the game mimics the one in the real world.
\item The two characters in the Glight game can move large spheres and place bombs near them. The vector nature of the bomb force can be explored by placing both bombs on (a) one side of a sphere and (b) on opposite sides. If done with reasonable care, the difference in the resulting trajectories, shown in Fig.~\ref{bombsaway},  demonstrates that forces of the same magnitude can have different effects depending on their direction. Thus, force is shown to be a vector quantity. Students can do further exploration by using a variety of bomb locations and match the outcome with their expectation.
\item Students construct a sled that can move on the glass track, one end of which is now terminated with a wall.  When a piston attached to the sled pushes on the wall, the sled accelerates away due to the reaction force exerted by the wall. By using a pendulum attached to the sled as an accelerometer, as shown in Fig.~\ref{lbpsled}, the effect of the reaction can be observed.
 This vividly illustrates the third law.
\end{enumerate}

\subsection{Motion under uniform gravity}
Traditional topics such as projectile motion under uniform gravity can be studied through in-game measurements.

\paragraph{Activities:}
\begin{enumerate}
\item  Students build a cannon in LBP and study the effects of the launch angle and initial speed on the range and height of a projectile. The value of acceleration due to gravity used in the game is measured from the motion of the projectile.  
\item Students build a pendulum in LBP. The period of the pendulum is used to measure the acceleration due to gravity and the value obtained can be cross-checked with that obtained from projectile motion.
 \end{enumerate}


\subsection{Translation and rotation of rigid bodies}
\begin{figure}
\includegraphics[scale=0.3,angle=-90]{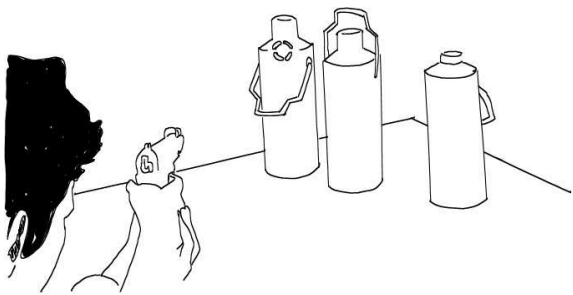}
\caption{
\label{shoottheflasks}
Experimenting with rigid body motion in the UnC2 game.  The translational and rotational motion of a flask depends on where it is shot. This is controlled by the player using the on-screen reticle (white circle near the top of the red flask).
}
\end{figure}

The sophistication of physics simulation in UnC2 allows students to see realistic dynamics, such as the rotation of rigid bodies, which are hard to come by in simpler games.  An example  is shown in Fig.~\ref{shoottheflasks}. Here the character is shown aiming a gun at a thermos flask. 

\noindent {\it Activity:} Shoot the flasks at various locations and observe their translational and rotational motion. Hitting near the top sends the flask into a clockwise rotation (seen from the right of the shooter) while hitting the bottom leads to anti-clockwise rotation. Hitting the middle leads to very little rotation. 

This example introduces students to the concept of torque and how the variation of torque leads to different types of rotations. The student learns that the same force vector applied at different points of an extended body can result in different rotational motion.  
\subsection{Violations of physics}
A physics simulation, by definition, has flaws.  However, these flaws can also be used to create teachable moments. For example, in some scenes of UnC2, a character can be made to stand on thin air at the edge of a cliff or a body hits the floor and partially sinks into it. Similarly, in the Glight game, balls launched by exploding bombs sometime ricochet off invisible boundaries.
\paragraph{Activity:} Students spot violations of physics and discuss why such effects do not occur in the real world. For example, in the real world bodies are kept from moving into each other by normal forces acting between them.
 On the other hand, games use a mathematical algorithm for collision detection which can suffer from errors.

Another teachable moment in UnC2 is the violation of momentum conservation when people get shot. Whereas a straightforward calculation shows that firing a machine gun into a body should impart enough momentum to throw the body backwards, the opposite happens in many instances. 
Students are made aware that 
  the computational load of complex animations is often reduced in games by exploiting human tolerance for error perception 
  (Reitsma \& Pollard, 2003; Yeh, Reinman, Patel \& Faloutsos, 2009). For example, falling human bodies in UnC2 are not
simulated faithfully but seem to be motion capture sequences blended with some ragdoll physics. 

Learning about violations of physics in games could enable students to better appreciate the profundity of physical laws in 
the real world. This can be emphasized further in laboratory experiments by having students contrast real world and game physics.

\section{Discussion}
\label{discussion}

  As most practicing physicists know, making diagrams is an integral part of solving  problems in physics.  
  In this respect, one of the most important benefits of games for students  is the
 development of visual thinking and  visualization 
 skills. The activities in Sec~\ref{paths}, for example, aid students in developing 
three-dimensional visual thinking, which is 
 hard to achieve using flat and static diagrams.   This is also evident from the following comment by a student received on a survey carried out after the UTB course: ``I got to see first hand with the almost concrete ideas what we were learning ... The games helped me see more clearly what we were trying to imagine, things like acceleration vectors.''
  
Most of the measurement activities we have described would be difficult to accomplish in the real world due to physical limitations imposed by classrooms and student numbers. Although these foundational activities are best learnt through hands-on activities,
 a traditional classroom setting limits most students to be passive observers.  
 With access to a game console, either in the classroom or 
 outside, every student can  practice these basic concepts. 
 We saw evidence of this in the UTB course when one of the students brought her own game (``God of War"), to the classroom to discuss the physics in that game. She had noted that some of  the moves allowed for the main character violated the first law of Newtonian mechanics. Thus, games allow this kind of self-motivated learning to continue outside the classroom.

In some cases, such as measuring coordinates of points on a path in the SWS game, it is difficult to obtain precise values. 
It may be feasible, however, to use video tracking software on recordings of game play. An example is provided by Allain  for the game ``Angry Birds" (Allain, 2010). Although this is a two-dimensional game whereas SWS involves motion in three dimensions, the basic idea may work in the latter.

We find that commercial games of today are of limited use for topics other than Newtonian mechanics.  Other physics topics, such as electrostatics, are seen rarely or simulated with very low fidelity. However, educational games, such as ``Supercharged!" developed by Squire \etal (2004), can fill this gap. Thus, 
by developing an eclectic mix of commercial and educational games, the educational community can 
 bring the established benefits of gaming (Honey \& Hilton, 2011) into the classroom in a cost-effective way.

\ack
The course on which this paper is based was funded by the Dean, Dr. M. Bouniaev, of the College of Science, Mathematics and Technology at the University of Texas at Brownsville. The chair of the Physics department at UTB, Dr. S. Mukherjee, and her staff, Ms. Wiley and Ms. Mireles, played a key role in supporting the logistics of this course. 

\section*{References}
\begin{harvard}
\item Allain R 2010 URL://www.wired.com/wiredscience/2010/10/physics-of-angry-birds/ (last accessed on Apr 1, 2011).
\item Boeing A \& Brunl T 2007 Evaluation of real-time physics simulation systems {\it  GRAPHITE '07: Proc.  5th international conference on Computer graphics and interactive techniques in Australia and Southeast Asia}, pp. 281. (New York: ACM)
\item Honey M A \& Hilton M (ed) 2011 {\it  Learning science through computer games and simulations} (Washington, D.C: The National Academies Press).
\item Price C B 2008 Learning physics with the Unreal Tournament engine {\it Phys. Educ.} {\bf 43} 291.
\item Reitsma P S A \& Pollard N S 2003 Perceptual metric for character animation: sensitivity to errors in ballistic motion {\it  Proc. SIGGRAPH '03} vol 22(3) (New York:ACM) pp.537-542
\item Squire K, Barnett M, Grant J M \& Higginbotham T 2004 Electromagnetism supercharged!: learning physics with digital simulation games {\it ICLS '04: Proc. 6th Int. Conf. on Learning Sciences} International Society of the Learning Sciences,  pp 513-520
\item Yeh T Y, Reinman G, Patel S J \& Faloutsos P 2009 Fool me twice: Exploring and exploiting error tolerance in physics-based animation, {\it J. ACM Trans. Graph. (TOG)} {\bf 29}(1) 5.
\end{harvard}
\end{document}